\documentclass[twocolumn,showpacs,preprintnumbers]{revtex4-1}
\usepackage{amsfonts}
\usepackage{amsmath}
\usepackage{graphicx}
\usepackage{dcolumn}
\usepackage{bm}
\usepackage{epstopdf}
\usepackage{hyperref}
\usepackage{subfigure}
\usepackage[dvipsnames]{xcolor}

\begin{document}

\preprint{}
\title{A method for computing driving and detuning beam coupling impedances of an asymmetric cavity using eigenmode simulations}
\author{Sergey Arsenyev}
\email{sergey.arsenyev@cern.ch}
\author{Benoit Salvant}
\affiliation{CERN, Geneva, Switzerland}
\date{\today }

\begin{abstract}
We propose a method for numerical calculation of driving and detuning transverse beam coupling impedances of an asymmetric cavity.
The method relies on eigenmode simulations and can be viewed as an alternative to time-domain wakefield simulations.
A similar procedure is well-established for symmetric cavities, and this paper extends it to the case of an asymmetric cavity.
The method is benchmarked with time-domain wakefield simulations and its practical implementation is discussed.
\end{abstract}

\maketitle

\section{Introduction}
\label{sec:intro}

In particle accelerators, the interaction of the particle beam with its surrounding is a source of coherent beam instabilities that may limit the performance of the machine.
This interaction can be described by the wake function $W(t)$ in the time domain, or, alternatively, by the impedance function $Z(\omega)$ in the frequency domain ($t$ is the time and $\omega$ is the angular frequency).
In this paper, we focus specifically on the transverse impedance $Z_{\perp}(\omega)$ - a function that describes the transverse kick received by the trailing charge due to the fields excited by the driving charge.
Let us call the transverse direction of interest $x$ and the transverse impedance of interest $Z_x(\omega)$.

The total transverse impedance $Z_x(\omega)$ (measured in $\Omega$) is a function of both the driving charge offset $(x_d,y_d)$ and the trailing charge offset $(x_t,y_t)$, with $y$ being the second transverse coordinate.
To the first order, the expansion of the total impedance around a point of interest $(x_0, y_0)$ can be written as
\begin{equation}
\begin{split}
Z_x(\omega, x_d, x_t, y_d, y_t) = c_1 + (x_d - x_0) c_2 \\
+ (x_t - x_0) c_3 + (y_d - y_0) c_4 + (y_t - y_0) c_5,
\end{split}
\end{equation}
with the coefficients of expansion $c_{1-5}$ first introduced in~\cite{Heifets} and~\cite{Tsutsui}.
The coefficients describing the dependencies on the driving charge offset $x_d$ and the trailing charge offset $x_t$ are defined as the driving and the detuning impedances (measured in $\Omega / m$), respectively:
\begin{equation}
\begin{split}
Z_x^{driv}(\omega, x_0) \equiv c_2 = \frac{\partial}{\partial x_d} Z_x(\omega, x_d, x_t)|_{x_d=x_t=x_0}, \\
Z_x^{det}(\omega, x_0) \equiv c_3 = \frac{\partial}{\partial x_t} Z_x(\omega, x_d, x_t)|_{x_d=x_t=x_0}.
\end{split}
\label{Z_driv_def}
\end{equation}
where we omit the irrelevant $y$-dependence.

The driving and the detuning transverse impedances are also sometimes referred to as the dipolar and the quadrupolar (as in, for example,~\cite{Mounet_thesis}).
This is due to the similarity of their effect on the trailing charge to that of a dipole magnet and of a quadrupole magnet: a constant kick independent of $x_t$, and a kick proportional to $x_t$, respectively.
This terminology, however, has nothing to do with the concept of dipolar and quadrupolar RF resonator modes, classified by the number of angular variations of the fields (see, for example,~\cite{Padamsee}, p.41).
To avoid this confusion, the ``driving-detuning'' terminology is used here instead.

We define the sum of $Z^{driv}(\omega, x_0) $ and $Z^{det}(\omega, x_0) $ as $Z^\Sigma (\omega, x_0)$, also referred to as the ``generalized term" in~\cite{Carlo}.
It describes the dependence of the total impedance on the transverse offset in the case when $x_d = x_t$.
In certain cases, $Z^\Sigma (\omega, x_0)$ is easier to measure (a single wire measurement is sufficient~\cite{Hugo_Day}) and easier to numerically calculate (see section~\ref{sec:discussion} below).
However, $Z^\Sigma (\omega, x_0)$ by itself is not sufficient to study beam stability, as $Z^{driv}$ and $Z^{det}$ affect the beam in different ways, as was pointed out in~\cite{Danilov_Burov} and discussed in~(\cite{Benoit_thesis} pp.44--47).
For example, beam stability simulations with codes HEADTAIL~\cite{Headtail} and PyHEADTAIL~\cite{Pyheadtail} account for both $Z^{driv}$ and $Z^{det}$, while the codes like DELPHI~\cite{Delphi} and Nested Head-Tail~\cite{NHT} make use of only the driving impedance.
In either case, $Z^{driv}$ and $Z^{det}$ have to be separated.

In the past, a procedure for separate measurements of $Z^{driv}$ and $Z^{det}$ with the wire techniques has been established~\cite{Tsutsui, Hugo_Day}.
In parallel, numerical tools have been developed with the possibility to separately compute the impedances.
One of the most powerful numerical tools is the time-domain wakefield simulation (e.g. CST wakefield solver~\cite{CST}).
However, a time-domain calculation can be time-consuming and often there is a need for verification with an alternative numerical solver.
If an impedance is of the resistive-wall type, calculations based on the field-matching technique (code IW2D~\cite{IW2D}) or the transmission line technique (code TLWALL~\cite{Carlo}) can serve as this alternative.

In this paper, we focus not on the impedance of the resistive-wall type, but on the impedance of the geometric type, and, in particular, impedance due to electromagnetic modes trapped in a resonant cavity.
In this case, as well, an alternative to time-domain calculations exists.
Namely, the impedance is found through the sum of the eigenmode solutions, each one characterized by its angular frequency $\omega_{res}$, \emph{shunt} impedance $R$ and quality factor $Q$.
In this paper, we focus on this eigenmode approach and extend the range of applicability of the already existing method to the case of an asymmetric cavity.
We define the cavity to be asymmetric if it does not possess the mirror symmetry in the direction of interest (defined here as $x$) around the beam trajectory.
Note that a cavity symmetric around a point different from the beam trajectory will be considered asymmetric for our purpose.

In the symmetric case, $Z^{driv}$ and $Z^{det}$ are clearly separated, because each mode can have \emph{either} driving \emph{or} detuning non-zero impedance (a proof of this will be given in section~\ref{sec:discussion}).
The modes thus can be classified into ``driving modes" and ``detuning modes".
In both cases, the impedance of a given mode can be found with the transverse resonator formula~(\cite{Mounet_thesis}, p.72):
\begin{equation}
\begin{split}
Z_x^{sym~resonator}(\omega, x_0) = \frac{\omega_{res}}{\omega}\frac{(\omega_{res} / c)R_x(x_0)}{1+iQ \left( \frac{\omega}{\omega_{res}}-\frac{\omega_{res}}{\omega}\right)} \\
= \frac{c}{2 \omega}\frac{R_{||}''(x_0)}{1+iQ \left( \frac{\omega}{\omega_{res}}-\frac{\omega_{res}}{\omega}\right)}
\end{split}
\label{Z_sym_res}
\end{equation}
where $Z_x^{sym~resonator}\equiv Z_x^{driv}$ if the mode is driving, $Z_x^{sym~resonator}\equiv Z_x^{det}$ if the mode is detuning, $\omega_{res}$ is the angular resonance frequency, $c$ is the speed of light, $R_x$ and $R_{||}$ are the transverse and the longitudinal shunt impedances defined here according to the ``circuit definition''~(\cite{Padamsee}, p.47), both measured in $\Omega$:
\begin{equation}
\left( \frac{R}{Q}\right)_{||}(x)=\frac{|V_{||}(x)|^2}{2 \omega_{res} U},~\left( \frac{R}{Q}\right)_x (x)=\frac{|V_x(x)|^2}{2 \omega_{res} U}.
\label{RoverQ}
\end{equation}
Here $Q$ is the quality factor of the resonant mode, $V_{||, x}(x)$ is the complex longitudinal and transverse voltages in the cavity at the offset $x$, and $U$ is the energy stored in the mode.
An additional factor of $(\omega_{res} / c)$ was inserted by the authors in Eq.~(\ref{Z_sym_res}) since the original formula in reference~\cite{Mounet_thesis} assumes a different definition for the transverse shunt impedance (measured in $\Omega / m$). The transition from $R_x$ to $R_{||}''$ in Eq.~(\ref{Z_sym_res}) is made using the Panofsky-Wenzel theorem (see section~\ref{sec:derivation} for details).

The validity of the classification into driving modes and detuning modes disappears in the case of an asymmetric cavity, where each mode can possess both $Z^{driv}$ and $Z^{det}$, with any ratio between the two.
As will be shown below, Eq.~(\ref{Z_sym_res}) no longer gives $Z^{driv}$ or $Z^{det}$, but instead gives $Z^\Sigma$.
To the best knowledge of the authors, no rigorously derived expressions for $Z^{driv}$ and $Z^{det}$ existed for asymmetric cavities.
This paper aims to fill in this gap, thus enabling the use of the eigenmode method as an alternative to the time-domain method.
Despite the time required for the postprocessing of the eigenmode data, the eigenmode method in many cases can be advantageous to the time-domain calculation.
The problems of the time-domain method include resonant modes of very high Q-factors that take too long to decay, numerical noise due to the beam injection scheme and the field integration, and mesh constraints for most codes.
Even in the cases when the time-domain method can be applied, the eigenmode method still serves as a valuable independent benchmarking tool.

\section{Derivation}
\label{sec:derivation}
In this section, we will derive an analogue of the transverse resonator formula (Eq.~\ref{Z_sym_res}) that would work for the case of an asymmetric cavity.
In fact, Eq.~(\ref{Z_sym_res}) will be replaced with two formulas (one for $Z^{driv}$ and one for $Z^{det}$), as in the asymmetric case a mode can possess both the driving and the detuning impedances.
To do so, we will start (perhaps, counter-intuitively) with the \emph{longitudinal} resonator formula that is valid regardless of the symmetry of the cavity.
For our purpose, we will expand the longitudinal resonator formula to the case of unequal offsets of the driving and the trailing charges.
Doing so will allow us to obtain the \emph{transverse} driving and detuning impedances by applying the Panofsky-Wenzel theorem.

The longitudinal resonator formula that works for the case $x_d=x_t$, is given by~(\cite{Mounet_thesis}, p. 72)
\begin{equation}
\begin{split}
Z_{||}^{resonator}(\omega, x_d=x_t=x_0)= \\
=\frac{R_{||}(x_0)}{1+iQ \left( \frac{\omega}{\omega_{res}}-\frac{\omega_{res}}{\omega}\right)} \equiv f(x_0) g(\omega),
\end{split}
\label{long_resonator}
\end{equation}
where we have separated the dependences on the offset and the frequency in the functions
\begin{equation}
\begin{split}
f(x) \equiv \left( \frac{R}{Q} \right)_{||} (x), \\
g(\omega) \equiv \frac{Q}{1+iQ \left( \frac{\omega}{\omega_{res}}-\frac{\omega_{res}}{\omega}\right)}.
\end{split}
\end{equation}

We now expand Eq.~(\ref{long_resonator}) to the more general case when the trailing charge does not necessarily follow the exact path of the driving charge.
In using Eq.~(\ref{long_resonator}) we have made an assumption that the electromagnetic fields in the cavity are given by the single resonant mode.
The time-behavior of the fields is dictated by the frequency and the $Q$-factor of the mode, and is, therefore, the same everywhere in the cavity.
This means that the choice of the integration line $x_t$ does not affect the frequency response $g(\omega)$, but only changes the amplitude of the response.
The amplitude depends now on both $x_d$ and $x_t$ through some function $F$:
\begin{equation}
Z_{||}^{resonator}(\omega, x_d, x_t)=F(x_d, x_t) g(\omega).
\label{expanded}
\end{equation}

To find the function $F$, we examine the longitudinal voltage kick received by the trailing charge $\Delta V(t)$.
This kick can be written as $\Delta V(t) = - q W_{||}$ by the definition of the wake function, which for the resonator impedance is given by~(\cite{Mounet_thesis}, p.72)
\begin{equation}
\begin{split}
W_{||}(t) = \omega_{res} F(x_d, x_t) e^{-\alpha t} \\
\times
\left( \mathrm{cos}(\bar{\omega}_{res} t) - \frac{\alpha}{\bar{\omega}_{res}} \mathrm{sin} (\bar{\omega}_{res} t)\right),
\end{split}
\end{equation}
where $\bar{\omega}_{res} = \omega_{res} \sqrt{1 - \frac{1}{4Q^2}}$, $\alpha = \frac{\omega_{res}}{2 Q}$, and $\left( \frac{R}{Q} \right)_{||}$ has been replaced with $F(x_d, x_t)$.
For us, the only important part of this wake function is the coefficient $\omega_{res} F(x_d, x_t)$ in front of the time-structure.
The time-structure is irrelevant for the purpose of this derivation and can be ignored by considering the voltage kick at a time much lower than the resonant period $T = 2 \pi \omega_{res}$
\begin{equation}
\Delta V(t \ll T) = - q \omega_{res} F(x_d, x_t).
\label{dv_1}
\end{equation}
This kick can also be written using the concept of the shunt impedance of the cavity.
Immediately after the passage of the driving charge, the cavity voltage is in the decelerating phase~(see \cite{Padamsee}, p.333), giving $\Delta V(t \ll T) = -|V_{||}|$.
The cavity voltage is related to the energy stored in the mode $U$ through the shunt impedance at the position of the trailing charge $x_t$ as $|V_{||}| = \sqrt{f(x_t) \times 2 \omega_{res} U}$.
In turn, $U$ is determined by the energy initially deposited in the cavity by the driving charge $U = k q^2$, where $k = \omega_{res} f(x_d) / 2$ is the loss factor ~(\cite{Vaccaro}, p.9) at the position of the driving charge $x_d$.
This gives
\begin{equation}
\Delta V(t \ll T) = - q \omega_{res} \sqrt{f(x_d) f(x_t)}
\label{dv_2}
\end{equation}
By comparing Eq.~(\ref{dv_1}) to Eq.~(\ref{dv_2}), the unknown function $F$ is found as $F(x_d, x_t) = \sqrt{f(x_d) f(x_t)}$.
Finally, the generalized longitudinal resonator formula is
\begin{equation}
Z_{||}^{resonator}(\omega, x_d,x_t) = \sqrt{f(x_d) f(x_t)} g(\omega).
\label{expanded2}
\end{equation}

We now apply Panofsky-Wenzel theorem (in the same way as in~\cite{Vaccaro}, p 23) to Eq.~(\ref{expanded2}) to get the total transverse impedance $Z_x(\omega, x_d, x_t)$
\begin{equation}
\begin{split}
Z_x(\omega, x_d, x_t) = \frac{c}{\omega} \frac{\partial}{\partial x_t} Z_{||}(\omega, x_d, x_t) \\
= \frac{c g(\omega)}{2 \omega} \sqrt{\frac{f(x_d)}{f(x_t)}} f'(x_t)
\end{split}
\end{equation}
We then find the driving and detuning impedances using their definitions~(\ref{Z_driv_def}) to get the final expressions
\begin{equation}
\begin{split}
Z_x^{driv}(\omega, x_0) = \frac{c g(\omega)}{4 \omega} \frac{{f'(x_0)}^2}{f(x_0)} \\
Z_x^{det}(\omega, x_0) = \frac{c g(\omega)}{4 \omega} \left( -\frac{{f'(x_0)}^2}{f(x_0)} + 2 f''(x_0) \right).
\end{split}
\label{Zdriv_det}
\end{equation}

\section{Discussion}
\label{sec:discussion}
First, it is interesting to notice that that the sum of the two equations~(\ref{Zdriv_det}) gives
\begin{equation}
Z_x^{\Sigma}(\omega, x_0) = \frac{c g(\omega)}{2 \omega} f''(x_0),
\end{equation}
which matches the symmetric cavity expression~Eq.~(\ref{Z_sym_res}).
This means that Eq.~(\ref{Z_sym_res}) when applied to the case of an asymmetric cavity, gives neither $Z^{driv}$ nor $Z^{det}$, but their sum.
As was stated in the introduction, finding $Z_x^{\Sigma}$ is easier as it only requires the first polynomial coefficient of the parabolic fit, and is not sensitive to the errors in $f$ and $f'$ (see below).

Second, the general~Eq.~(\ref{Zdriv_det}) can be applied to a completely symmetric cavity as a particular case.
One can infer from Eq.~(\ref{Zdriv_det}) that the modes in a symmetric cavity are either purely detuning or purely driving.
This distinction comes naturally as the two following cases: non-zero longitudinal impedance $f(x_0) \neq 0$, and zero longitudinal impedance $f(x_0) = 0$.
To show this, we first note that $f'(x_0) = 0$ due to the symmetry.
It means that in the case $f(x_0) \neq 0$ the driving impedance is equal to zero, and the detuning impedance becomes $Z_x^{det}(\omega, x_0) = \frac{c g(\omega)}{2 \omega} f''(x_0)$ confirming Eq.~(\ref{Z_sym_res}).
In the other case $f(x_0) = 0$ we have a $0/0$-type fraction, hence the limit has to be taken when approaching the point of symmetry $x_0$.
The $0/0$-type fraction is resolved by Taylor-expanding the function $f$ around $x=x_0$ as $f(x) = \frac{f''(x_0)}{2} (x-x_0)^2 + O((x-x_0)^3)$.
The two terms in the second equation of~(\ref{Zdriv_det}) cancel out, giving zero detuning impedance.
The driving impedance becomes $Z_x^{driv}(\omega, x_0) = \frac{c g(\omega)}{2 \omega} f''(x_0)$, again confirming Eq.~(\ref{Z_sym_res}).
To sum up, in a symmetric cavity, a mode with non-zero longitudinal impedance has no driving contribution, while a mode with zero longitudinal impedance has no detuning contribution.

\begin{figure}[!htb]
\centering
\includegraphics[width=70mm, angle=0]{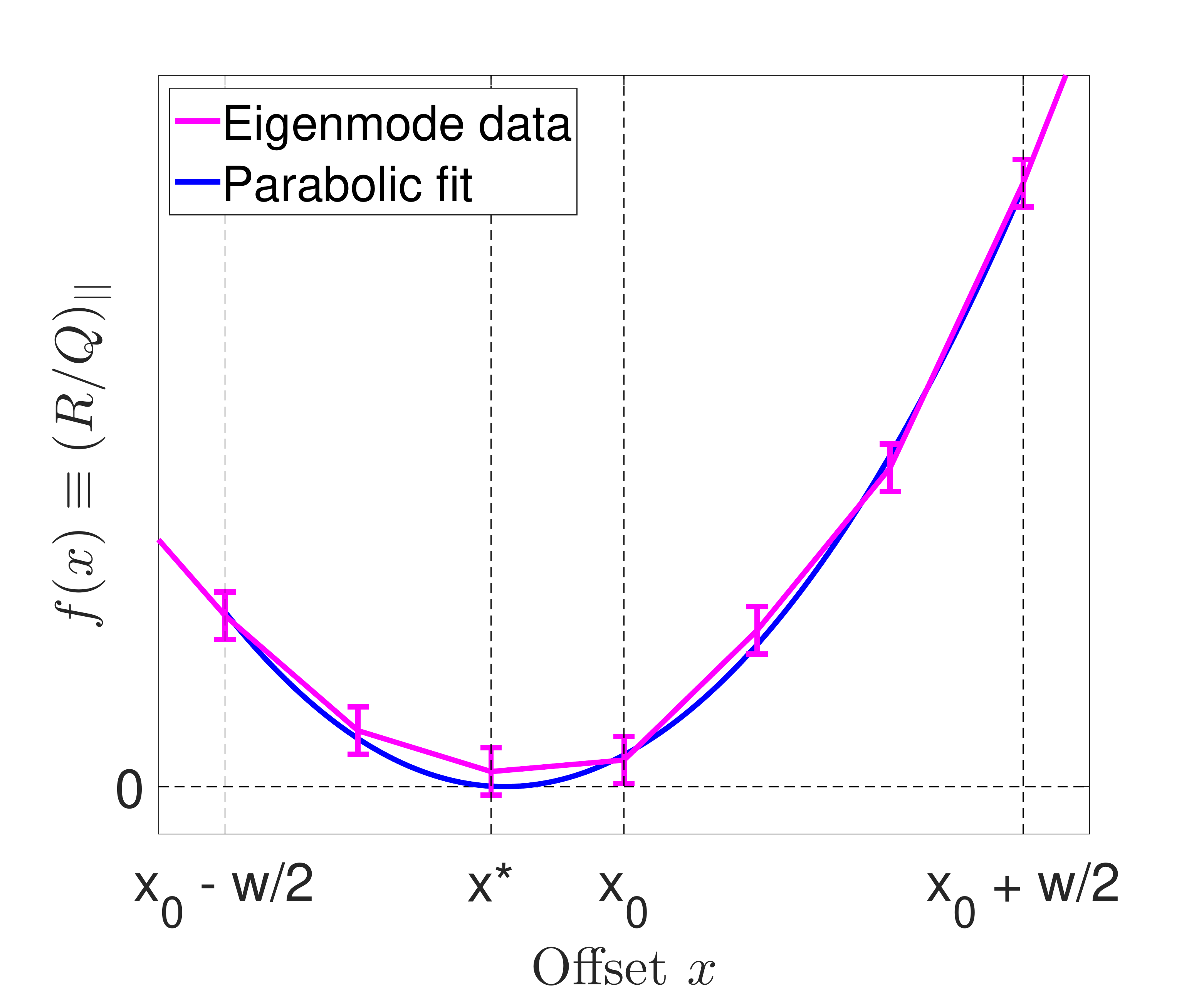}
\caption{An illustrative example of eigenmode data for a slightly asymmetric cavity. The fit window around the point of interest $x_0$ contains 7 data points (the purple line with the error bars). The value of $f$ falls below its error bar inside the fit window, and the tangent constraint is enforced to the fitting parabola (the blue line).}
\label{example}
\end{figure}

Third, a practical implementation of Eq.~(\ref{Zdriv_det}) deserves a separate discussion.
If $f'$ and $f''$ are calculated by numerical differentiation, even a small numerical error in $f$ will result in significant uncertainties.
A more practical solution is to do a parabolic fit to the data in some window $w$ around the point of interest $x_0$ (see Fig.~\ref{example}), and determine $f'$ and $f''$ from the corresponding coefficients.
In section~\ref{sec:brick}, the two ways will be compared by applying them to a simple cavity with and without artificially added computational noise.
Special attention should be taken when the method is applied to a symmetric or to a slightly asymmetric cavity (e.g. the only source of asymmetry is an HOM coupler).
In that case, driving modes have vanishingly low $f(x_0)$ and $f'(x_0)$, and blindly applying~Eq.~(\ref{Zdriv_det}) will result in diverging results for both $Z_x^{driv}$ and $Z_x^{det}$ due to the term $f'^2/f$.
The problem only applies to driving modes, as for detuning modes the term in question simply goes to zero since $f'(x_0) =0 $ and $f(x_0) \neq 0$.

One way to avoid this problem is to constrain the fitting parabola to go through zero (the ``tangent constraint''), thus resolving the $f'^2/f$ ratio.
This constraint can be triggered when the value of $f$ falls below the error bar of the eigenmode simulation within the fitting window, as shown in Fig.~\ref{example}.
The constraint is implemented by looking for the fit not in the general form $f = a x^2 + b x + c$, but in the form $f = a (x - x^*)^2$ with $x^*$ somewhere in the window $(x_0 - w/2) \leq x^* \leq (x_0 + w/2)$.
This solution allows to correctly determine $Z_x^{driv}$, however at the expense of losing the small non-zero value of $Z_x^{det}$ (plugging $f = a (x - x^*)^2$ in~Eq.~(\ref{Zdriv_det}) yields exactly zero detuning impedance).
If in fact there is a need to determine the low level of $Z_x^{det}$, the fitting window $w$ should be adjusted not to cover the tangent point $x^*$, with the step between the sampling points scaled down accordingly.

\section{Check for a simple cavity}
\label{sec:brick}
\begin{figure}[!htb]
\centering
\includegraphics[width=70mm]{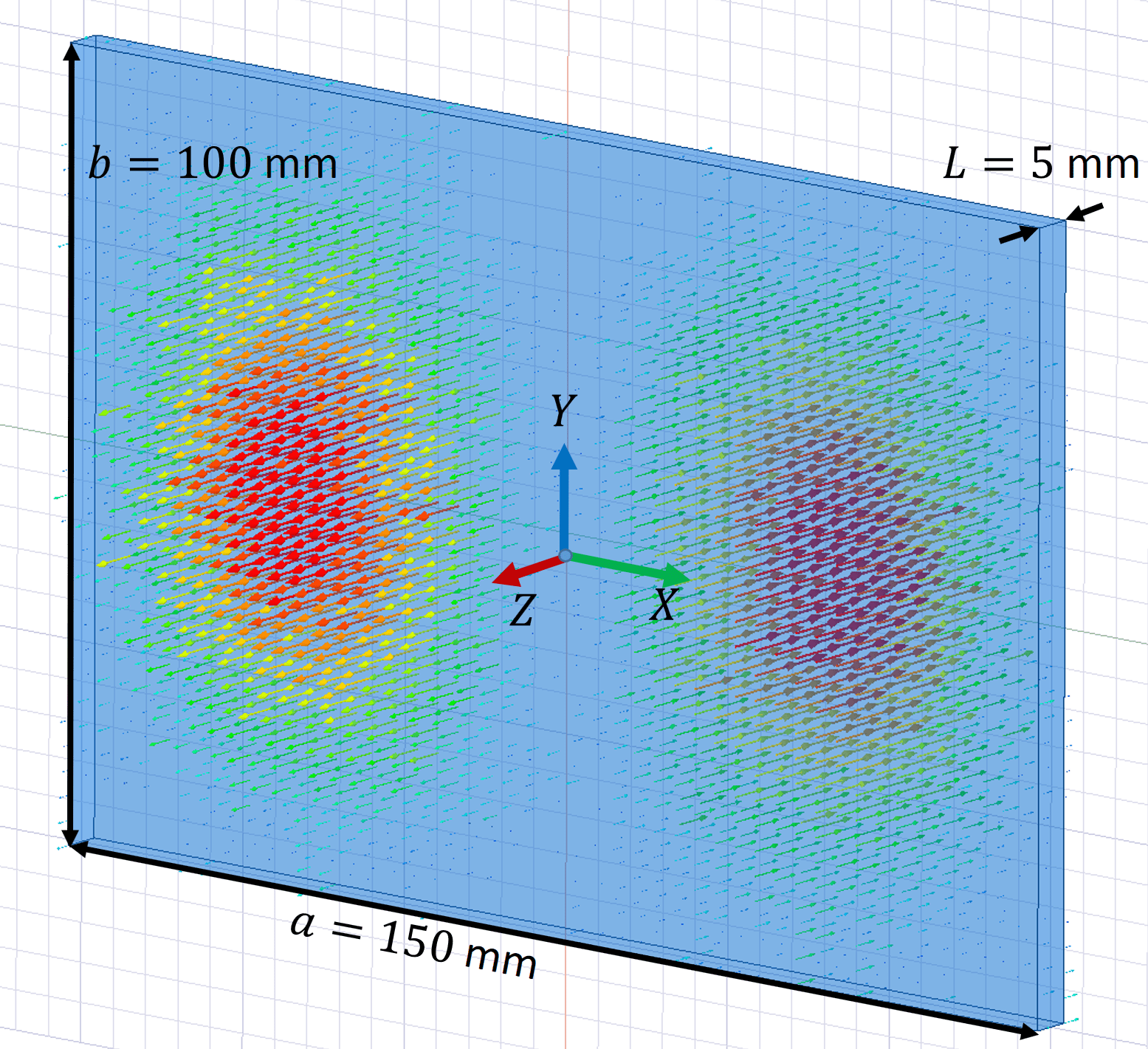}
\includegraphics[width=70mm]{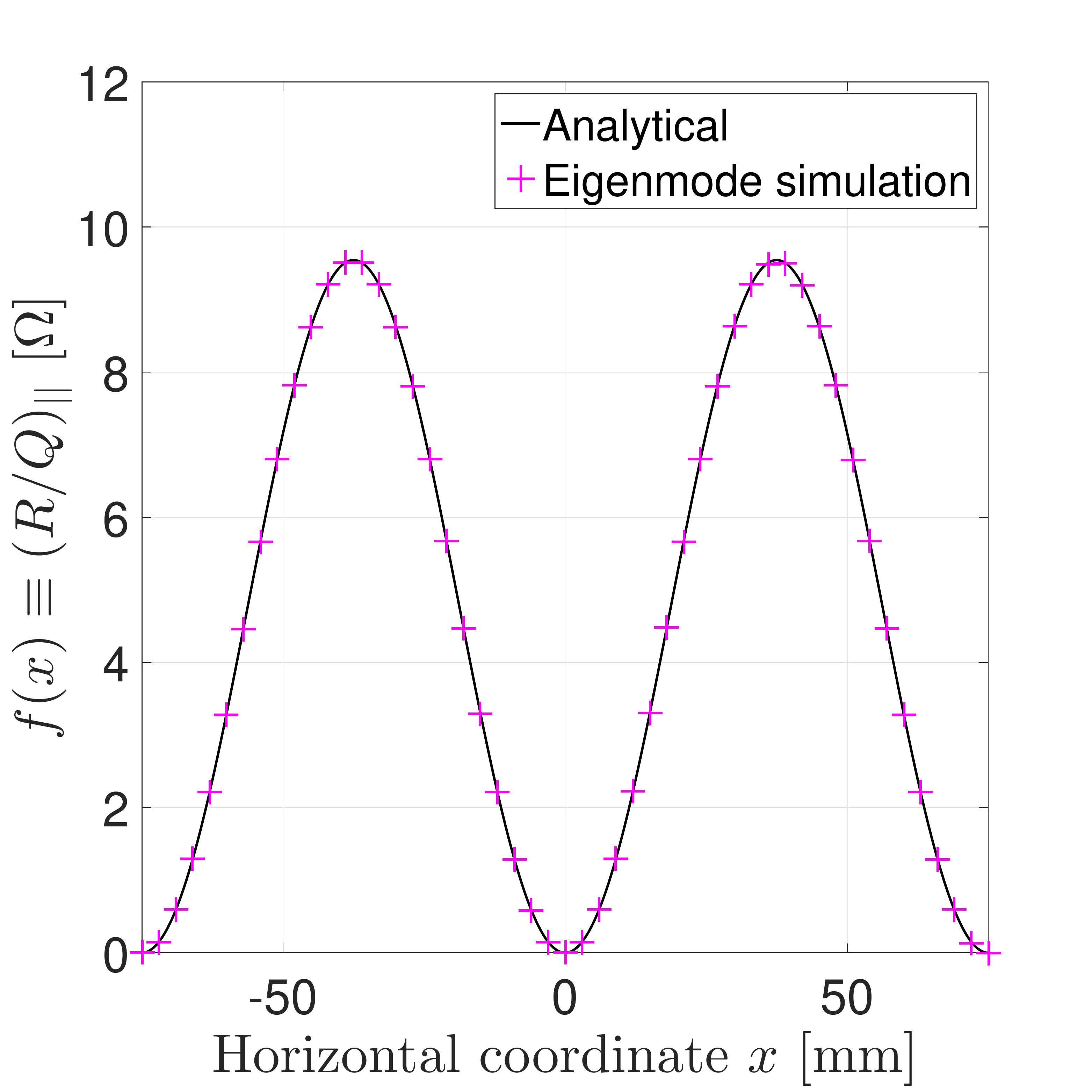}
\caption{Top: The considered example of a simple cavity (the vacuum region is shown in blue). The electric field of the chosen mode (TM210) is shown with arrows. Bottom: $f(x) \equiv (R/Q)_{||}$ of the selected mode as a function of the horizontal coordinate $x$, with the black line given by the analytical formula and the purple crosses given by the eigenmode simulation.}
\label{brick}
\end{figure}

As an example, we consider a $TM_{210}$ mode at the frequency of 2.5 GHz excited in a rectangular resonator shown in Fig.~\ref{brick} (top).
If the beamline is put in the center $x_0 = 0$, it corresponds to the case of a symmetric cavity, while a shifted beamline with $x_0 \neq 0$ corresponds to the asymmetric case.

Let us first examine the validity of Eq.~(\ref{Zdriv_det}) with no regard of numerical errors in the calculation of the shunt impedance.
For that, the simplicity of the considered geometry allows for an analytical derivation of $(R/Q)_{||}$.
The electric field in the selected mode is given by
\begin{equation}
\begin{split}
E_z (x, y, z, t) = A \mathrm{sin} \left( \frac{2 \pi x}{a} +\pi \right) \\
\times \mathrm{sin} \left( \frac{\pi y}{b} + \frac{\pi}{2} \right) e^{i \omega_{res} t}
\end{split}
\end{equation}
where $A$ is a constant, $a$ and $b$ are the dimensions of the cavity as shown in Fig.~\ref{brick}, and $\omega_{res}$ is the angular resonant frequency of the mode.
The voltage and the stored energy in the mode are computed as
\begin{equation}
\begin{split}
V_{||}=\int_0^L dz E_z(z, t=z/c), \\
U=L \int_{-a/2}^{a/2}dx \int_{-b/2}^{b/2}dy \frac{1}{2} \epsilon_0 E_z^2(t=0),
\end{split}
\end{equation}
where $L$ is the length of the cavity and $\epsilon_0$ is the vacuum permittivity.
Fixing the vertical coordinate $y$ to zero leads to the following expression for $f(x) = (R/Q)_{||}$
\begin{equation}
f(x) = M \mathrm{sin}^2 \left( \frac{2 \pi x}{a} \right),~M = \frac{4c^2 |e^{i \omega_{res} L/c} - 1|^2}{\omega_{res}^3 L \epsilon_0 ab}
\label{f_brick}
\end{equation}
which is shown in Fig.~\ref{brick} (bottom) as the solid black line.
Given Eq.~(\ref{Zdriv_det}), the driving and the detuning impedances for some offset $x_0$ become
\begin{equation}
\begin{split}
Z_x^{driv}(\omega, x_0) = \frac{c g(\omega)}{4 \omega} \times \frac{16 \pi^2}{a^2} M \mathrm{cos}^2 \left( \frac{2 \pi x_0}{a} \right),\\
Z_x^{det}(\omega, x_0) = -\frac{c g(\omega)}{4 \omega} \times \frac{16 \pi^2}{a^2} M \mathrm{sin}^2 \left( \frac{2 \pi x_0}{a} \right).
\end{split}
\label{quantities}
\end{equation}

\begin{figure}[!htb]
\centering
\includegraphics[width=90mm, angle=0]{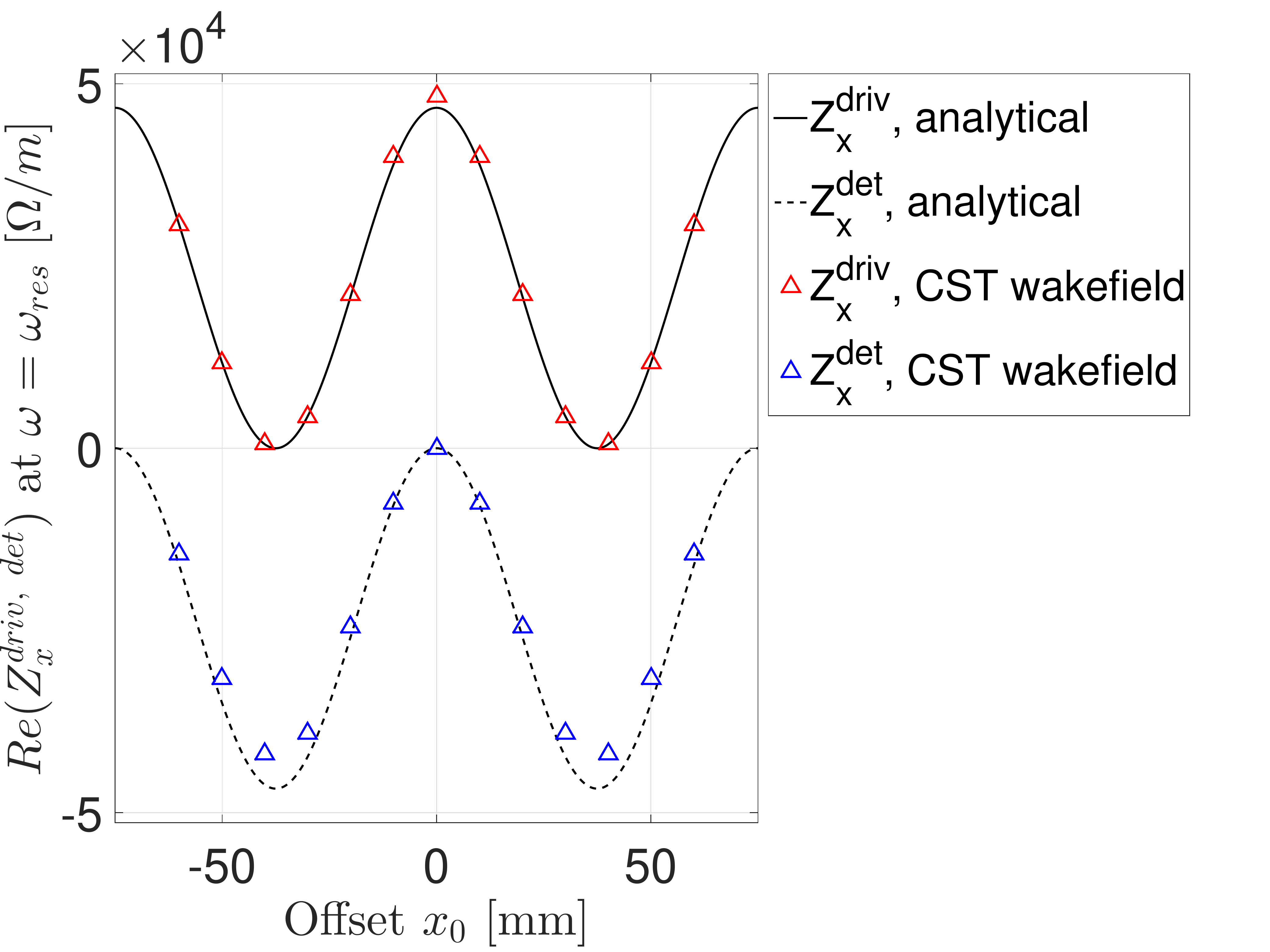}
\caption{Driving and detuning impedances obtained by the CST wakefield simulations, compared to the derived analytical expressions.}
\label{Z_comparison_CST}
\end{figure}

These quantities can be checked against time-domain wakefield simulations, for which the CST wakefield solver~\cite{CST} was used.
For this, the cavity was enclosed in the walls of lossy metal (conductivity $\sigma = 10^5~ S/m$) giving the considered mode the quality factor $Q = 146$.
For each of the few selected offsets $x_0$, a small opening was made in the walls around $x_0$ to allow for the passage of the beam.
The real part of the transverse impedance was observed for both the beam line and the integration line at $x_0$, and for small displacements of either line.
The variation of the height of the resonance impedance peak with the displacement of the beamline gave $Z_x^{driv}$, and the variation with the displacement of the integration line gave $Z_x^{det}$.
These quantities are plotted in Fig.~\ref{Z_comparison_CST} together with the analytical result (Eq.~(\ref{quantities})).
A generally good agreement is achieved for all offsets $x_0$ including the case of a completely symmetric cavity ($x_0=0$), which confirms the validity of the derived theory.
Note specifically that the point $x_0 = 0$ corresponds to the case $f(x_0) = 0,~f'(x_0) = 0$, and the point $x_0 = 37.5~mm$ corresponds to the case $f(x_0) \neq 0,~f'(x_0) = 0$ (Fig.~\ref{brick}, bottom).
The absence of the detuning impedance or the driving impedance at these points confirms the statement made in section~\ref{sec:discussion}.

So far we only showed that the derived theory gives correct results when we take the analytical expression for $f(x)$ as an input.
Another interesting question is how accurate would the results be if instead we took actual eigenmode simulations as an input.
To check this, we used an eigenmode solver (Ansys HFSS~\cite{HFSS} or CST Microwave Studio~\cite{CSTMS}) to find the values of $f$ at a series of offsets spaced by 3 mm (crosses in Fig.~\ref{brick}, bottom).
As expected for such a simple cavity, the data is in a very good agreement with the analytical result.
We then apply the two methods discussed in~\autoref{sec:discussion} to find $f'$ and $f''$ from the data.

\begin{figure}[!htb]
\centering
\includegraphics[width=90mm, angle=0]{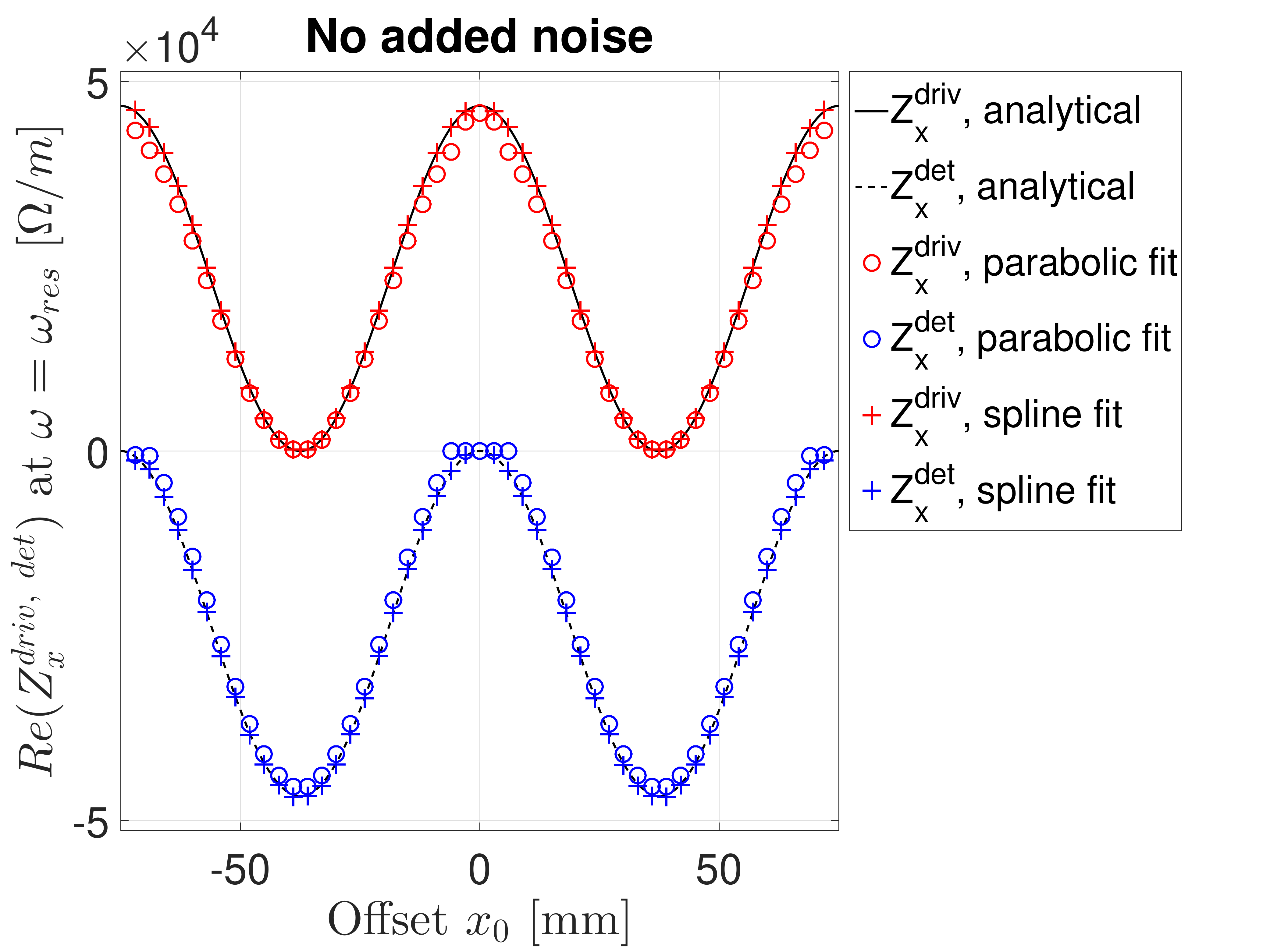}
\includegraphics[width=90mm, angle=0]{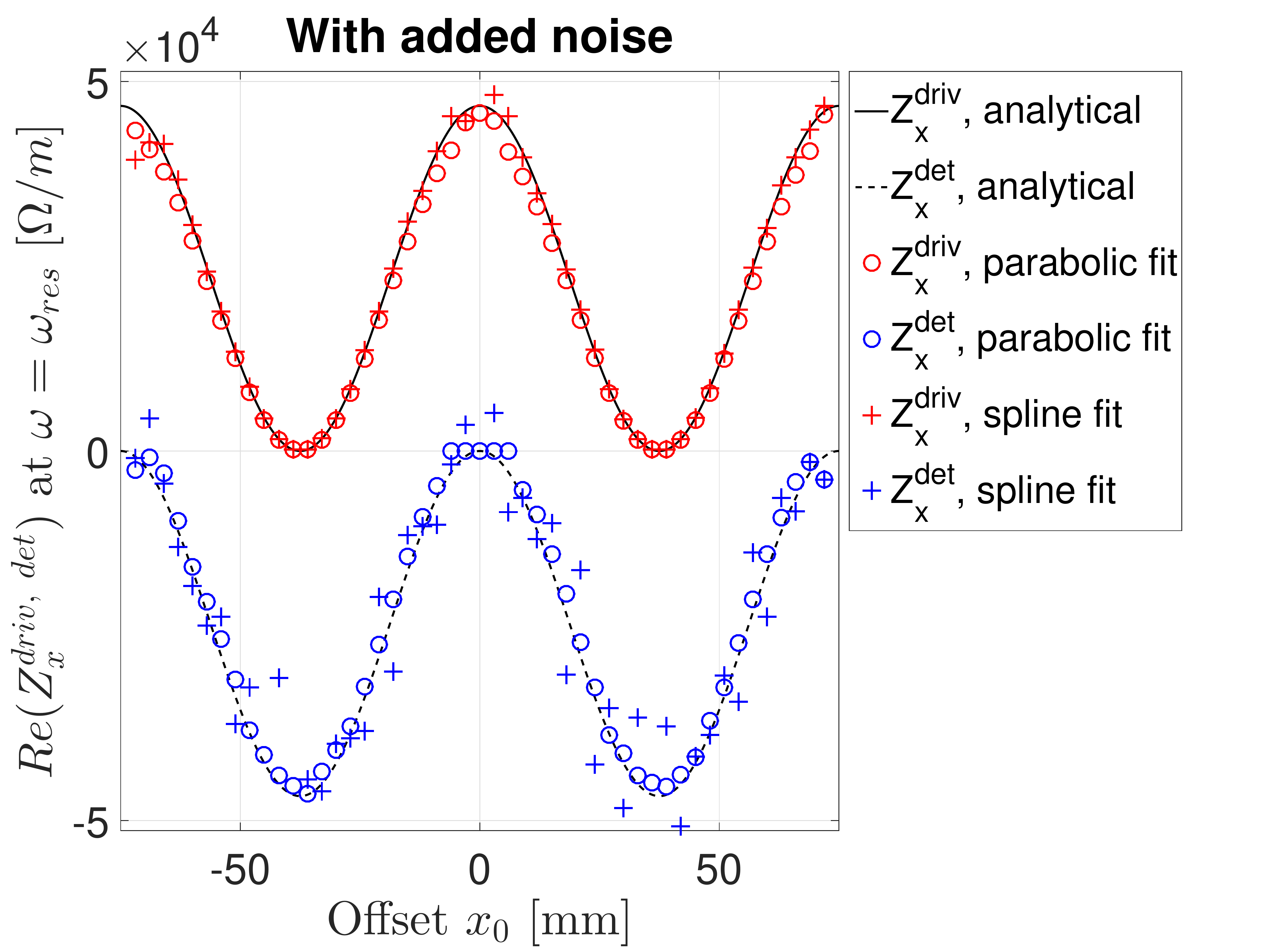}
\caption{Driving and detuning impedances obtained by the eigenmode method, compared to the analytical expressions. Top: the original eigenmode data is taken without added noise. Bottom: random errors of the amplitude of $0.025~\Omega$ were added to the eigenmode data.}
\label{Z_comparison_noise}
\end{figure}

The first method is to numerically differentiate $f(x)$ after spline-interpolating the data points to define the function everywhere between the points.
Since the eigenmode data for such a simple cavity is very precise, the resulting $Z^{driv}$ and $Z^{det}$ match nicely with the analytical result (crosses in Fig.~\ref{Z_comparison_noise}, top).
As was mentioned in~\autoref{sec:discussion}, for a completely symmetric cavity the numerical differentiation gives a diverging result, hence the missing crosses at $x_0 = 0$.

The second method is the parabolic fit.
For this, a window of $w=12$ mm was used when fitting the points, meaning that the fits are based on the values of $f$ at the point of interest, plus two additional points on each side.
The obtained $Z^{driv}$ and $Z^{det}$ are shown in Fig.~\ref{Z_comparison_noise} (top) as circles.
The tangent constraint was enforced for the parabolic fits that contain the central point ($x = 0$) in the fitting window.
This allowed for a correct estimation of $Z^{driv}$ even for the completely symmetric case (red circle at $x_0 = 0$), but at the expense of giving $Z^{det} = 0$ for the 5 central points, as was discussed in section~\ref{sec:discussion}.
The agreement of $Z^{driv}$ and $Z^{det}$ with the analytical result is not as good for the parabolic fit as it was for the numerical differentiation (circles vs crosses in Fig.~\ref{Z_comparison_noise}, top).
However, since the shape of the fitting function is constrained to be a parabola, this method is more robust when applied to eigenmode data with larger error bars.
To illustrate this, we artificially added random errors of relatively small amplitude ($0.025~\Omega$) to the data points in Fig.~\ref{brick} (bottom).
We then applied both methods to the new ``noisy" data, and obtained the impedances shown in Fig.~\ref{Z_comparison_noise} (bottom).
The results of the parabolic fit are less distorted by the added noise, and this method is therefore preferred for complex geometries.

\section{Check for a realistic cavity}
\begin{figure}[!htb]
\centering
\includegraphics[width=70mm]{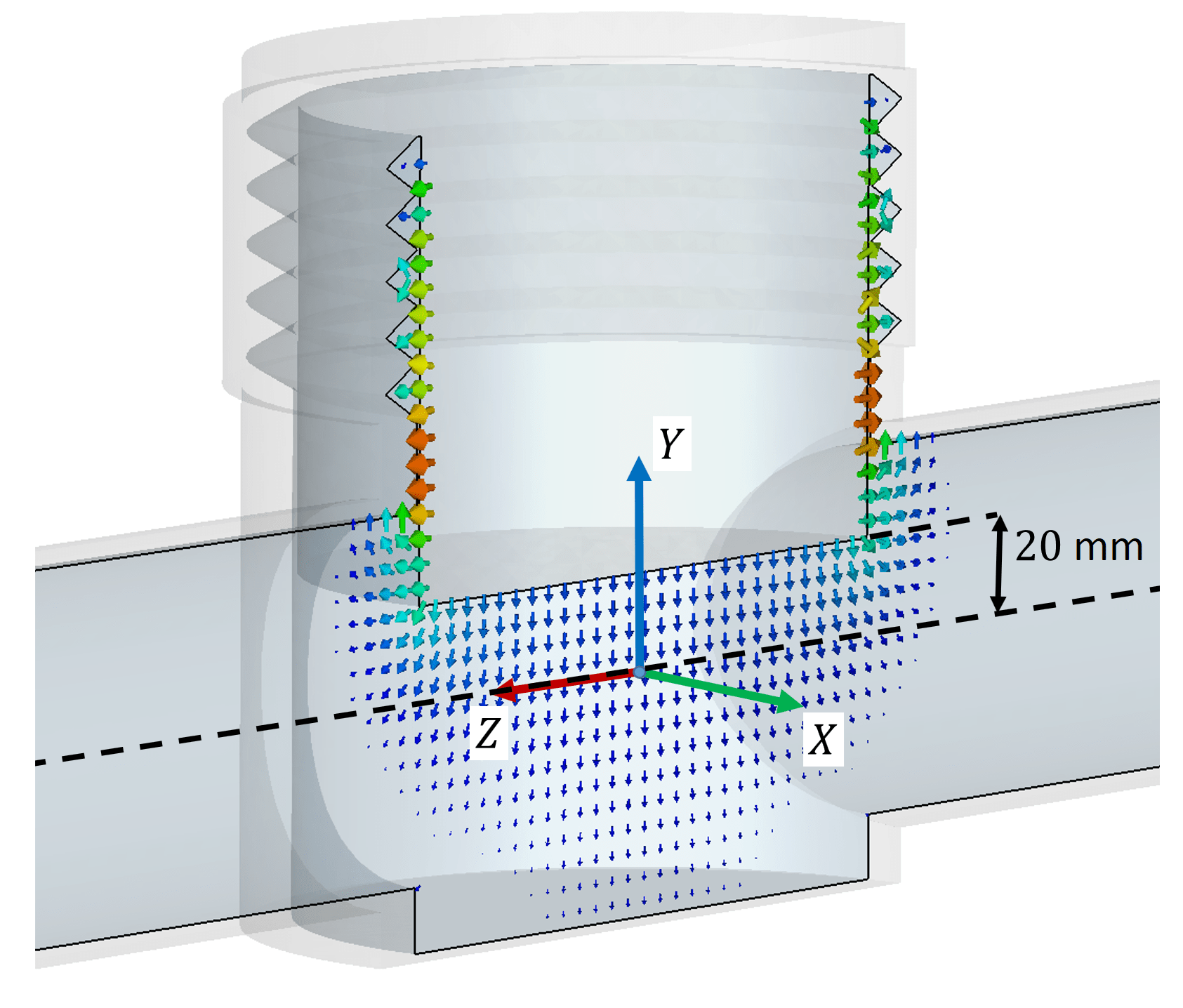}
\includegraphics[width=70mm]{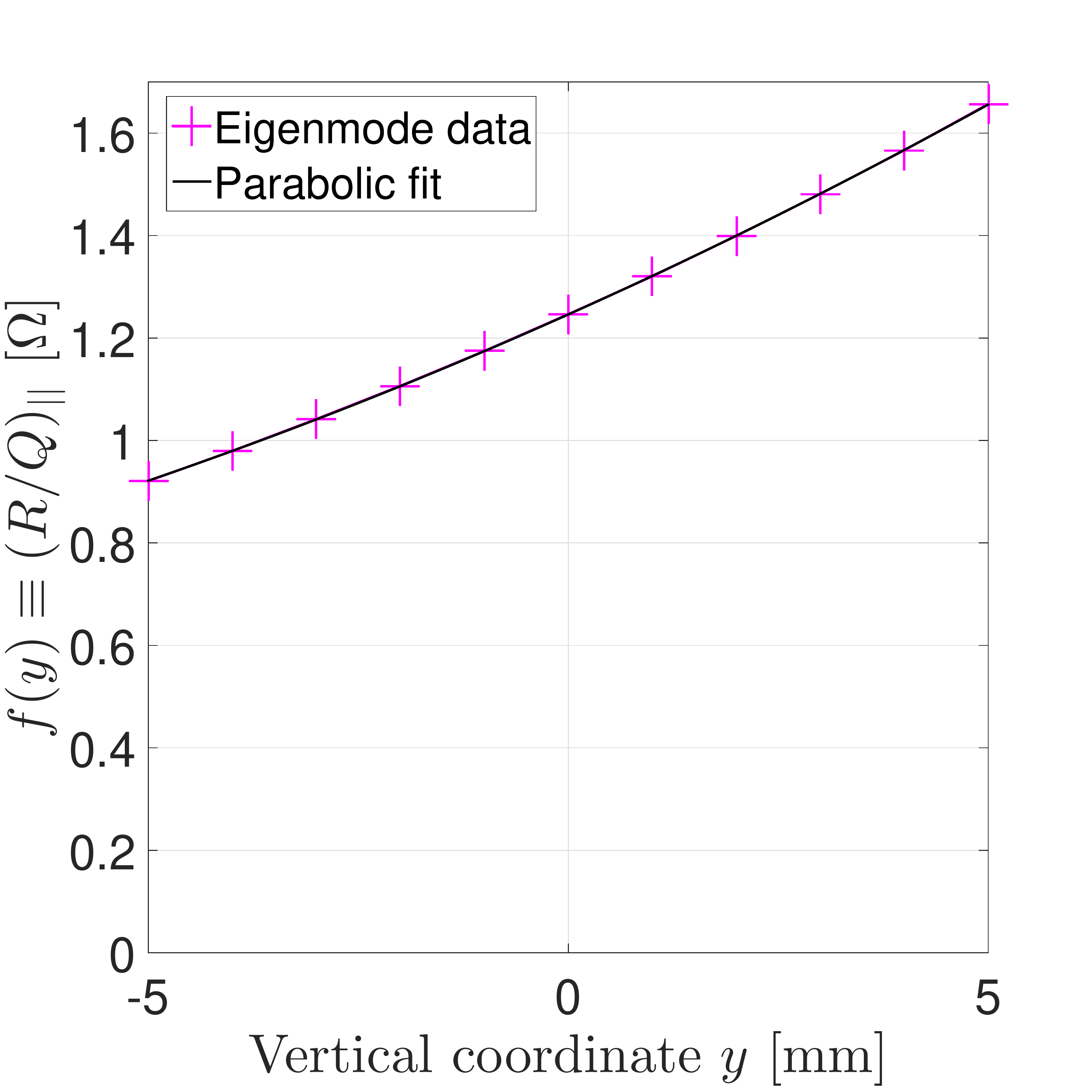}
\caption{Top: The considered roman pot cavity (coutesy: Nicola Minafra). The vacuum region is shown in grey. No ferrite absorbers are considered. The electric field of the chosen mode is shown with arrows. Bottom: $f(y) \equiv (R/Q)_{||}$ of the selected mode as a function of the vertical coordinate $y$.}
\label{roman_pot}
\end{figure}

As an example of an asymmetric cavity with complex geometry, we chose the roman pot structure shown in Fig.~\ref{roman_pot} (top).
The roman pot is an experimental technique for the detection of forward protons from elastic or diffractive scattering.
Detectors are placed inside a secondary vacuum vessel, called pot, and moved into the primary vacuum of the machine through vacuum bellows~\cite{roman_pot}.
For this example, we set the pot to be 20 mm away from the beam, resulting in a strongly asymmetric geometry in the vertical direction $y$.
The following analysis was therefore done for the vertical impedance $Z_y$.

\begin{figure}[!htb]
\centering
\includegraphics[width=90mm, angle=0]{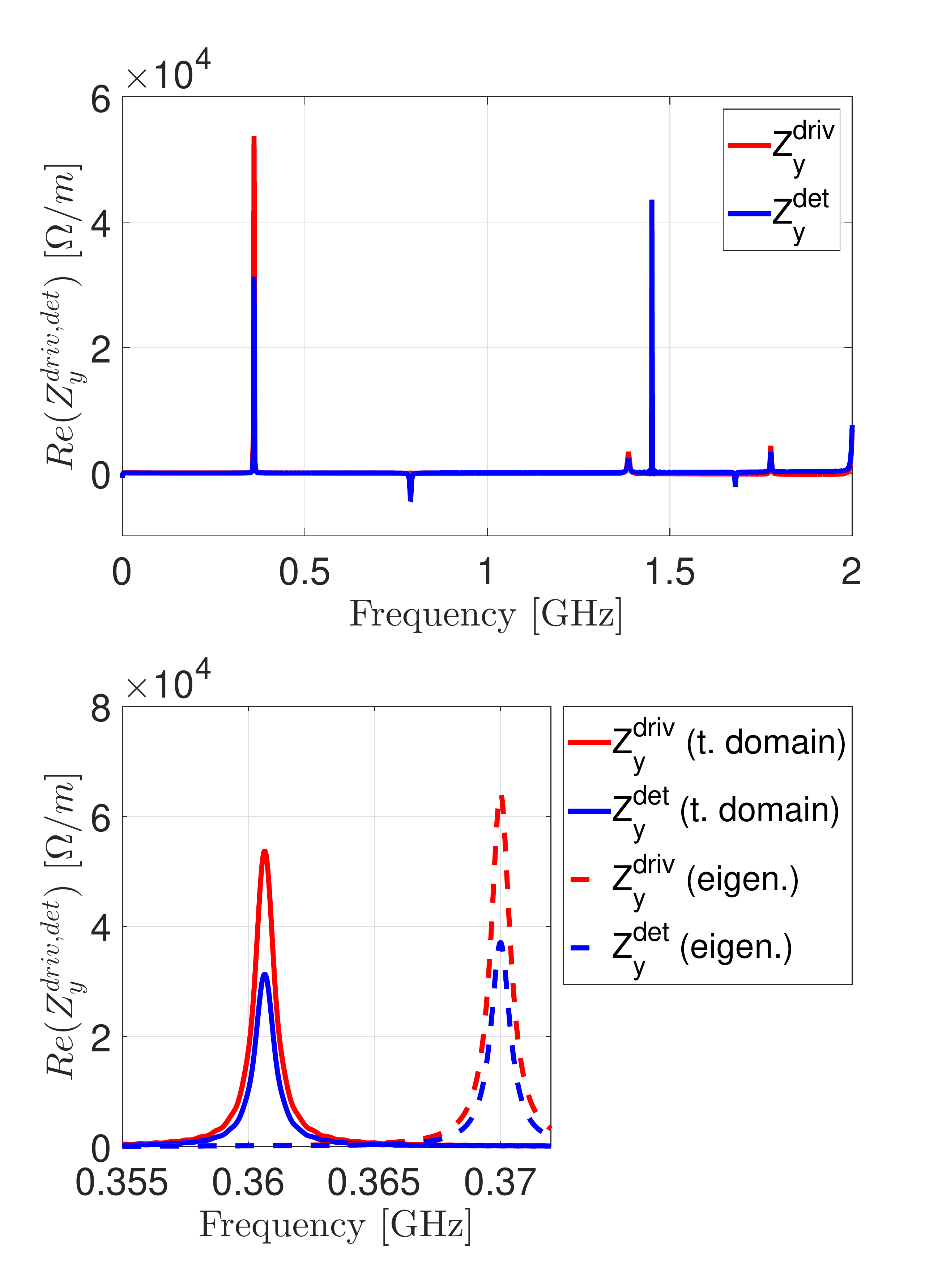}
\caption{Top: Wakefield calculation of the vertical transverse impedance in the roman pot structure. Bottom: A zoom-in on the most prominent peak, and a comparison to the eigenmode simulation.}
\label{roman_pot_wakefield}
\end{figure}

First, a traditional time-domain calculation was done using the CST wakefield solver in the same way as in~\autoref{sec:brick}, with the resulting driving and detuning impedances plotted in Fig.~\ref{roman_pot_wakefield} (top).
For comparison with an eigenmode analysis we chose the mode at the frequency of 360 MHz, corresponding to the most prominent peak in the impedance spectrum. 
The impedances at this resonant frequency are listed in Table~\ref{roman_pot_table}.
The chosen mode is primarily excited in the bellows region (see Fig.~\ref{roman_pot}, top).
The mode is coupled to the beam through the narrow gap between the pot and the outer walls.
The narrow gap requires sophisticated meshing and makes the mode a good candidate to test the eigenmode method.

Then, we performed the eigenmode analysis for the integration line offsets ranging from $-5$ to $5$ mm with a step of 1 mm.
The longitudinal shunt impedance and the parabolic fit are shown in Fig.~\ref{roman_pot} (bottom).
The resulting driving and detuning impedances are listed in Table~\ref{roman_pot_table} together with the resonant frequency and the Q-factor estimated with the eigenmode solver.
The corresponding resonant peak is also plotted against the wakefield data in Fig.~\ref{roman_pot_wakefield} (bottom).
Since the resonant frequencies obtained by the two methods are 10 MHz off (an error due to the complex geometry and the different mesh types), the eigenmode peak appears shifted but otherwise resembles the wakefield one for both $Z_y^{driv}$ and $Z_y^{det}$.

\begin{table}[!htb]
\centering
\caption{The driving and the detuning impedances of the roman pot cavity, calculated with the eigenmode method and with the time-domain wakefield simulations.}
\begin{tabular}{lcccr}
& $f_{res}$ & $Q$ & $Re(Z_y^{driv}(f_{res}))$ & $Re(Z_y^{det}(f_{res}))$ \\
\hline
Eigenmode & 370 MHz & 462 & $6.43\times10^4~\Omega/m$ & $3.71\times10^4~\Omega/m$ \\
T. domain & 360 MHz & 405 & $5.54\times10^4~\Omega/m$ & $3.14\times10^4~\Omega/m$ \\
\end{tabular}
\label{roman_pot_table}
\end{table}

In the absolute values, the two methods agree within 15\%.
Most of this difference can be explained by the different Q-factors in the eigenmode and the wakefield solvers.
Indeed, in the wakefield solver, the Q-factor deduced from the width of the resonance peak is lower than the eigenmode value (Table~\ref{roman_pot_table}).
This source of error can be canceled out if we use the Q-factor from the wakefield solver in the eigenmode formula.
Then, the difference between the two methods decreases to only 4\%.
As a conclusion, in this case of a realistic structure, the developed method provides the correct decomposition of detuning and driving impedances of the main resonant mode.

\section{Conclusions}
We developed a method to separately compute the driving (dipolar) and the detuning (quadrupolar) geometrical transverse impedances of an asymmetric cavity.
As an input, the method takes eigenmode data for a chosen cavity mode: the longitudinal shunt impedance $(R/Q)_{||}$ for several transverse offsets, and the mode's frequency and Q-factor.
The method was benchmarked for the case of a simple cavity (displaced rectangular resonator) and a realistic cavity (roman pot), showing a good agreement with time-domain calculations in both cases.
For practical implementation of the method, we also investigated how numerical noise in the input eigenmode data affects the end results.
It was concluded that a parabolic fit of the data is preferred when the method is applied to realistic cavities.

\section{Acknowledgements}
This work was supported by the European Union's Horizon 2020 research and innovation programme under grant No 654305.
We also thank Carlo Zannini, Agnieszka Chmielinska, and Elias M\'{e}tral for the useful discussions and Nicola Minafra for the choice of the benchmarking geometry.

\end{document}